\documentclass[sn-mathphys,Numbered]{sn-jnl}

\usepackage{graphicx}%
\usepackage{multirow}%
\usepackage{amsmath,amssymb,amsfonts}%
\usepackage{amsthm}%
\usepackage{mathrsfs}%
\usepackage[title]{appendix}%
\usepackage{xcolor}%
\usepackage{textcomp}%
\usepackage{manyfoot}%
\usepackage{booktabs}%
\usepackage{algorithm}%
\usepackage{fix-cm}
\usepackage{algorithmicx}%
\usepackage{algpseudocode}%
\usepackage{listings}%
\makeatletter
\newcommand*{\rom}[1]{\expandafter\@slowromancap\romannumeral #1@}
\makeatother

\theoremstyle{thmstyleone}%
\theoremstyle{thmstyletwo}%

\theoremstyle{thmstylethree}%

\begin{document}

\title[Article Title]{Singly poled thin film lithium niobate waveguide as a tunable source of photon pairs across telecom band}

\author{\fnm{Muskan} \sur{Arora}}\email{idz218420@iitd.ac.in}

\author{\fnm{Jyoti} \sur{Arya}}\email{jyotiaryaformal@gmail.com}

\author{\fnm{Pranav} \sur{Chokkara}}\email{idz218513@iitd.ac.in}

\author*{\fnm{Jasleen} \sur{Lugani}}\email{jasleen@sense.iitd.ac.in}

\affil{\orgdiv{Centre of Sensors, Instrumentation and Cyber Physical System Engineering (SeNSE)}, \orgname{Indian Institute of Technology Delhi}, \orgaddress{\street{Hauz Khas}, \city{New Delhi}, \postcode{110016}, \state{State}, \country{India}}}

\abstract{Spontaneous parametric down conversion (SPDC), especially in non-linear waveguides, serves as an important process to generate quantum states of light with desired properties. In this work, we report on a design of a strongly dispersive, singly poled thin film lithium niobate (TFLN) waveguide geometry which acts as a convertible source of photon pairs across telecom band with tunable spectral properties. Through our simulations, we demonstrate that by using this optimized waveguide geometry, two completely different yet desirable type II phase-matched SPDC processes are enabled using a single poling period. One process generates spectrally correlated non-degenerate photon pairs with one photon at 1310 nm (telecom O band) and the other at 1550 nm (telecom C band). The second SPDC process results in spectrally uncorrelated photon pairs in telecom C band at 1533 nm and 1567 nm respectively.We attribute this versatility of TFLN waveguide to its strong dispersion properties and make a comparative study with the existing weakly dispersive waveguide platforms.We believe that such a versatile source of photon pairs will serve as an important ingredient in various quantum optical tasks which require photons at different telecom bands and desired spectral properties.}




\maketitle

\section{Introduction}
Photon pair sources based on non-linear optical processes constitute one of the most fundamental and integral building blocks in various quantum optics experiments \cite{o2009photonic,wang2020integrated}.
  Several material platforms, especially in integrated architecture, have been explored for building efficient sources of quantum light with the desired spectral, spatial and temporal properties \cite{francesconi2023chip,chen2023generation,koefoed2023high}. Thin film lithium niobate  (TFLN) is one such platform which has attracted a lot of attention lately, not only for generating photon pairs but also for potentially manipulating them \cite{saravi2021lithium,vazimali2022applications,lomonte2020chip,colangelo2020superconducting}. Benefiting from the versatile material non-linear optical properties of lithium niobate and structuring of nano-scale waveguides, TFLN offers several advantages, most important of which is its high refractive index contrast and dispersion engineering capability, which helps in tailoring the spectral properties of the generated quantum light \cite{kumar2022group,briggs2021simultaneous}. 
  
TFLN has recently been explored to build several quantum light sources including high-quality broadband entangled photon pairs\cite{zhao2020high,duan2020generation,javid2021ultrabroadband}, spectrally pure photons \cite{kumar2022group} and squeezed light\cite{chen2022ultra}. These sources are based on a second-order non-linear optical process known as spontaneous parametric down-conversion (SPDC) \cite{kurtsiefer2001generation} which occurs when both energy and momentum of the interacting photons are conserved. However, due to material dispersion, momentum conservation is normally difficult to achieve \cite{boyd2008nonlinear}. One method to address this issue is through quasi-phase matching (QPM) technique, wherein, the direction of the instantaneous polarization in the interacting region is periodically reversed \cite{boyd2008nonlinear} with a certain period. This overall generates spatial frequencies in k-space which compensates for the phase mismatch. Usually, for enabling an SPDC process in a non-linear medium, a specific poling period is required to generate signal-idler pair at the desired wavelengths. In quantum optics, both telecom O and telecom C bands constitute highly desirable wavelength ranges, especially for communication tasks involving fiber networks\cite{lugani2020spectrally,martin2009integrated}. Thus, generating photon pairs at these wavelengths is of extreme interest. However, if a non-linear waveguide or crystal is designed and poled to generate photons in one of these telecom bands, the same poling period may not enable another SPDC process which produces photons in the other band as the respective poling period is very different.This is especially the case with the weakly dispersive bulk non-linear crystals \cite{anwar2021entangled,lee2016polarization} as well as conventional microscale waveguides including traditional lithium niobate ridge structures \cite{kumar2020postselection} and periodically poled potassium titanyl phosphate (PPKTP) waveguides \cite{meier2021comparison}. In such cases, enabling multiple desirable SPDC processes in the same waveguide or crystal demands special domain engineering \cite{sun2020microstructure} of the interaction region. This on the one hand is difficult to fabricate and on the other hand reduces the efficiency of the respective SPDC processes. Another solution would be to appropriately tailor the dispersion properties of the non-linear medium, wherein waveguide dispersion plays a crucial role.
In this work, we address this issue and explore the dispersion profile of a TFLN-based waveguide to simultaneously satisfy phase matching conditions of two different SPDC processes resulting in photons across  telecom O and C band using the same poling period. We observe that TFLN waveguides, due to their high refractive index contrast and small cross-section give rise to a very strong dispersion profile, which enables multiple desired SPDC processes using the same poling period.
In sections below, we detail and report on one such singly poled TFLN waveguide geometry that specifically switches between two desired SPDC processes giving rise to photon pairs in telecom O and C bands with different spectral properties, by just tuning the pump wavelength.

\section{Dispersion analysis}

 Dispersion is the key property which determines whether or not a given SPDC process is phase matched in a non-linear medium. In addition, it also governs the spectral properties of the generated bi-photon state. In waveguides, dispersion is not only determined by the constituting material but also its cross-sectional dimensions and geometry. While material dispersion is fixed, waveguide dispersion can be tailored to get the desired bi-photon state \cite{arora2024avoiding,kumar2022group}. In this context, one of the most important feature of TFLN waveguides is its strong waveguide dispersion, which allows for the possibility of enabling very different SPDC processes simultaneously using a single poling period. To understand this better, we discuss the two main conditions required to realize an SPDC process, namely, the energy conservation and momentum conservation. Energy conservation condition is given by
\begin{equation} 
\omega_{p} = \omega_{s} + \omega_{i},  
\end{equation}
where, $\omega_{p}$, $\omega_{s}$ and $\omega_{i}$ represent the frequencies of pump, signal, and idler photons involved in a SPDC process. In terms of wavelengths, Eq. (1) can be written as
\begin{equation} 
\frac{1}{\lambda_{p}} = \frac{1}{\lambda_{s}} + \frac{1}{\lambda_{i}}
\end{equation}
where, $\lambda_{p}$, $\lambda_{s}$, and  $\lambda_{i}$ are the wavelengths of the pump, signal, and idler photons respectively.
The second necessary condition for an SPDC process is the momentum conservation 
or the phase matching condition, which is
\begin{equation}
  k_p =  k_s + k_i
\end{equation}
where \textit{$k_{p(s,i)}$} is the propagation constant of pump (signal, idler) photons and is  related to its effective refractive index, \textit{$n_{p(s,i)}$} as   \textit{$k_{p(s,i)}$} =  $\frac{2 \pi n_{p(s,i)}}{\lambda_{p(s,i)}}$.
Usually, due to the material dispersion, phase matching is not automatically obeyed, i.e.  $\textit{$\Delta k$} =  k_p - k_s - k_i \neq 0$. To compensate for this phase mismatch (\textit{$\Delta k$}), the ferroelectric domains in the interaction region are periodically reversed. This effectively generates a spatial frequency $K$ which compensates for the phase mismatch.  $K$ is given by
 \begin{equation}
 \textit{K}  =  \frac{ 2\pi}{\Lambda}  = \textit{$\Delta k$}
 \end{equation}
where \textit{$\Lambda$} is the poling period required to phase match the given SPDC process.
In terms of effective refractive indices of pump ($n_{p}$), signal ($n_{s}$) and idler ($n_{i}$) photons, Eq.(4) can be written as 
\begin{equation}
 \textit{$\Delta k$} = 2\pi(\frac{n_{p}}{\lambda_{p}} - \frac{n_{s}}{\lambda_{s}} - \frac{n_{i}}{\lambda_{i}}) =  \frac{ 2\pi}{\Lambda}  
 \end{equation} 
Using energy conservation from Eq. (2) and substituting for  $\lambda_{p}$, Eq. (5)  reduces to
 \begin{equation}
 \textit{$\Delta k$} =2\pi( \frac{n_{p}-n_{s}}{\lambda_{s}} + \frac{n_{p}-n_{i}}{\lambda_{i}})=   \frac{ 2\pi}{\Lambda}
 \end{equation} 
 or in short,
 \begin{equation}
 \textit{$\Delta k$} =2\pi\sum_{r=s,i}\frac{n_{p}-n_{r}}{\lambda_{r}} = \frac{ 2\pi}{\Lambda}
 \end{equation}
 where subscript 'r' sums over signal and idler indices. It can be seen from Eq. (7) that the expression of phase mismatch (\textit{$\Delta k$})  
 involves the term $\frac{n_{p}-n_{r}}{\lambda_{r}}$ summed over signal and idler indices. This term can be understood as the ratio of the difference between effective refractive indices of pump and signal (or idler) to the signal (idler) wavelength. It is to be noted that all those SPDC processes (with the corresponding pump, signal and idler wavelengths) which carry the same phase mismatch ($\Delta k$), are enabled using the same period ($\Lambda$) in that waveguide. As we see in Eq.(7), this is mainly governed by the dispersion of the waveguide. 
Now, for weakly dispersive waveguides, the effective refractive index of a mode changes very gradually with wavelength. In this case, the numerator (${n_{p}-n_{r}}$) of Eq. (7) does not vary much from one SPDC process to another, however, the denominator  ($\lambda_r$) varies significantly if the respective downconverted photon pairs have very different sets of wavelengths. The phase mismatch ($\Delta k$) for such SPDC processes (as we detail below) are very different and thus can not be compensated using the same poling period \cite{shukla2021polarization}. However, the situation is different for strongly dispersive waveguides, in which the effective refractive index varies significantly with wavelength. As we move from one SPDC process to another, the appreciable changes in numerator term (${n_{p}-n_{r}}$) of (Eq.7) may compensate for the overall changes in the downconverted wavelengths in denominator term ($\lambda_{r}$). This can give rise to the possibility of satisfying multiple SPDC processes across different telecom bands using the same poling period. 
To understand this better, we consider three very different SPDC processes and compare different waveguide platforms to enable them. Focussing on type II phase-matched SPDC processes in which downconverted signal and idler photons have orthogonal polarizations, we consider the following three SPDC processes that generate signal-idler photons in telecom O and C bands: \\
 
  1) \hspace{0.5cm}  655 nm $\rightarrow$  \hspace{1cm}        1310 nm              \hspace{1cm} +   \hspace{1cm}  1310 nm  \\
  \hspace*{1.4cm}  (Pump)      \hspace{1.4cm}        (Signal)           \hspace{2.6cm}               (Idler) \\
 
2) \hspace{0.5cm} 775 nm $\rightarrow$ \hspace{1cm} 1550 nm \hspace{1cm} +   \hspace{1cm}   1550 nm  \\
 \hspace*{1.4cm}  (Pump)      \hspace{1.4cm}        (Signal)           \hspace{2.6cm}               (Idler) \\ 
 
3) \hspace{0.5cm} 710 nm $\rightarrow$ \hspace{1cm} 1310 nm \hspace{1cm} +   \hspace{1cm}   1550 nm  \\
 \hspace*{1.4cm}  (Pump)      \hspace{1.4cm}        (Signal)           \hspace{2.6cm}               (Idler) \\ \\
We now compare weakly and strongly dispersive waveguides by considering three waveguide systems with different cross-sectional areas and refractive index contrast ($\Delta n$). These include:  (1) a conventional microscale PPLN ridge waveguide (cross-sectional area $\approx$ 24 $\mu m^2$, $\Delta n$ = 0.8) \cite{kumar2020postselection}, (2) a PPKTP channel waveguide (cross-sectional area $\approx$ 16 $\mu m^2$, $\Delta n$ = 0.01)\cite{meier2021comparison}, and (3) a nanoscale TFLN rib waveguide (cross-sectional area $\approx$ 0.64 $\mu m^2$,$\Delta n$ = 0.8)\cite{arora2024avoiding}. Figure 1 shows the respective dispersion profiles (variation of effective refractive indices  ($n_{eff}$) of fundamental TE and TM modes with wavelengths) of these waveguide geometries. The dispersion of the PPLN ridge (red and purple solid lines) and PPKTP waveguides (doted blue and orange lines) has been computed using references (\cite{kumar2020postselection} and \cite{meier2021comparison}) while the dispersion of TFLN waveguide is taken from our previous work \cite{arora2024avoiding}, and is represented by dashed (pink and green) lines in Fig. 1. 

\begin{figure} [ht]
\begin{center}
         {%
    \includegraphics[width= 9 cm , height = 6 cm]{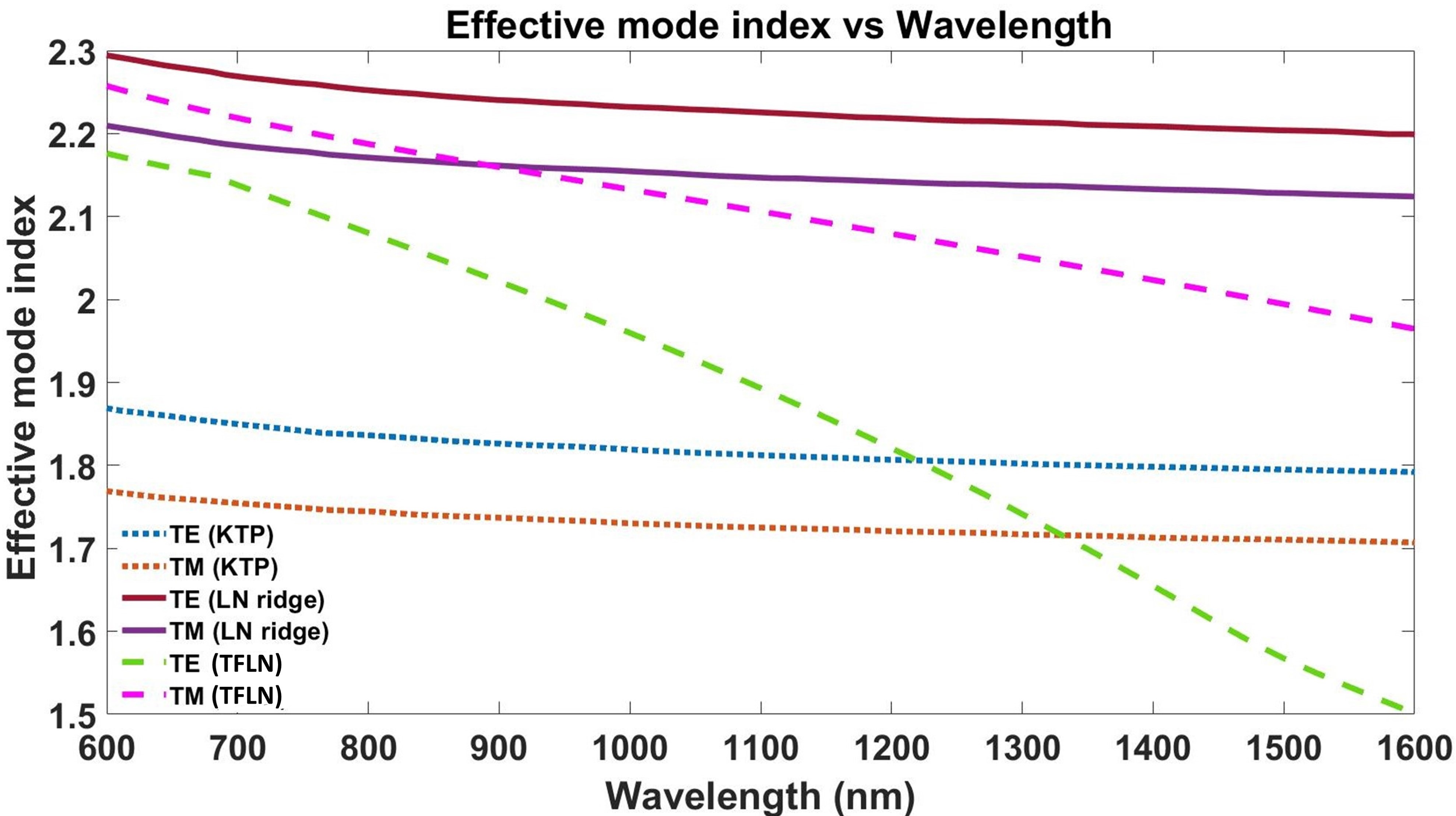}}%
        \caption{Variation of effective refractive index (vs wavelength) of (TE and  TM) modes of a PPLN ridge waveguide \cite{kumar2020postselection} represented by solid lines (Red and purple), a PPKTP channel waveguide \cite{meier2021comparison} represented by doted lines (blue and orange) and a TFLN rib waveguide \cite{arora2024avoiding} represented by dashed lines (green and pink).}
        \vspace{-1em}
        \label{FIG. 1}
         \end{center}
        \end{figure} 
 As can be seen from Fig 1, the slope of the dispersion curves for TFLN waveguide for both TE and TM polarization is steeper than the other two cases. This indicates much stronger dispersion for TFLN waveguides, which is attributed to its smaller cross-section (0.64 $\mu m^2$ ) \cite{arora2024avoiding} and larger refractive index contrast ($\Delta n$ $\approx$  0.8). Due to this, any change in the wavelength significantly changes the refractive index of the mode.
To see how this enables multiple desired SPDC processes using the same poling period, we compute the respective phase mismatches ($\Delta k$) and poling periods ($\Lambda$) using Eq. (7) for the three waveguide configurations by using the dispersion profiles of Fig.1. This is tabulated in Table 1.
From our calculations, we observe that for TFLN waveguide, the values of phase mismatches  ($\Delta k$) for the two of the above SPDC processes ((2) and (3)) are very close to each other (although different from $\Delta k$ of SPDC process (1)). This suggests the potential of enabling the respective processes using the same poling period. In contrast, for the other two waveguide configurations (PPKTP and PPLN), the respective phase mismatches ($\Delta k$) for all the  SPDC processes are very different and thus can not be simultaneously phase-matched using the same poling period.\\  
 
\begin{table}[ht]
 \caption{Comparison between different waveguide systems}
\vspace{2mm} 
\begin{tabular}{|l|l|l|l|}
\hline
                        & \textbf{PPLN \cite{kumar2020postselection}} & \textbf{PPKTP \cite{meier2021comparison}} & \textbf{TFLN \cite{arora2024avoiding}} \\ \hline
Dimensions (W $\times$ H) (in $\mu m^2$)   &     (6 $\times$ 4.04)  &   (4 $\times$ 4) & (0.8 $\times$ 0.8)     \\ \hline
\begin{tabular}[c]{@{}l@{}} Refractive index contrast ($\Delta n$) \\ 
 \end{tabular}   &    ~ 0.8           &   ~ 0.01     &  ~ 0.8     \\ \hline
 \begin{tabular}[c]{@{}l@{}} Phase mismatch (in $\mu m^{-1}$) 
 \\(for type II SPDC) \\ 1) 655 nm $\rightarrow$          1310 nm               +    1310 nm  \\ \hspace{0.5 cm}{\fontsize{6pt}{0.5pt}\selectfont (pump)}  \hspace{0.5 cm}                    {\fontsize{6pt}{0.5pt}\selectfont (signal)} \hspace{0.5 cm}   {\fontsize{6pt}{0.5pt}\selectfont (idler)}   \\ 2) 775 nm $\rightarrow$          1550 nm               +    1550 nm  \\ \hspace{0.5 cm}{\fontsize{6pt}{0.5pt}\selectfont (pump)}  \hspace{0.5 cm}                    {\fontsize{6pt}{0.5pt}\selectfont (signal)} \hspace{0.5 cm}   {\fontsize{6pt}{0.5pt}\selectfont (idler)} \\  3) 710 nm $\rightarrow$          1310 nm               +    1550 nm  \\ \hspace{0.5 cm}{\fontsize{6pt}{0.5pt}\selectfont (pump)}  \hspace{0.5 cm}                    {\fontsize{6pt}{0.5pt}\selectfont (signal)} \hspace{0.5 cm}   {\fontsize{6pt}{0.5pt}\selectfont (idler)}  \end{tabular}   &    \begin{tabular}[c]{@{}l@{}} \\ 1.012 \\ \\0.749 \\ \\ 0.825   \end{tabular}              &  \begin{tabular}[c]{@{}l@{}}  \\  0.025 \\ \\  0.060 \\ \\ 0.076 \end{tabular}     &  \begin{tabular}[c]{@{}l@{}} \\ 1.933 \\ \\ 1.7749 \\ \\ 1.7799  \end{tabular}    \\ \hline
 \begin{tabular}[c]{@{}l@{}} Poling period (in $\mu m$) 
 \\(for type II SPDC) \\  1) 655 nm $\rightarrow$          1310 nm               +    1310 nm  \\  \hspace{0.5 cm}{\fontsize{6pt}{0.5pt}\selectfont (pump)}  \hspace{0.5 cm}                    {\fontsize{6pt}{0.5pt}\selectfont (signal)} \hspace{0.5 cm}   {\fontsize{6pt}{0.5pt}\selectfont (idler)}   \\ 2) 775 nm $\rightarrow$          1550 nm               +    1550 nm  \\  \hspace{0.5 cm}{\fontsize{6pt}{0.5pt}\selectfont (pump)}  \hspace{0.5 cm}                    {\fontsize{6pt}{0.5pt}\selectfont (signal)} \hspace{0.5 cm}   {\fontsize{6pt}{0.5pt}\selectfont (idler)} \\  3) 710 nm $\rightarrow$          1310 nm               +    1550 nm  \\  \hspace{0.5 cm}{\fontsize{6pt}{0.5pt}\selectfont (pump)}  \hspace{0.5 cm}                    {\fontsize{6pt}{0.5pt}\selectfont (signal)} \hspace{0.5 cm}   {\fontsize{6pt}{0.5pt}\selectfont (idler)}  \end{tabular}   &    \begin{tabular}[c]{@{}l@{}} \\  6.21 \\ \\ 8.39  \\ \\  7.61  \end{tabular}              &  \begin{tabular}[c]{@{}l@{}}   \\  247 \\ \\   104 \\ \\   83.20 \end{tabular}     &  \begin{tabular}[c]{@{}l@{}} \\  3.25 \\ \\ 3.54 \\ \\ 3.53 \end{tabular}    \\ \hline
\end{tabular}
\end{table}

Thus, from our analysis and the comparative study above, we observe that the waveguides with stronger dispersion are more potent in satisfying multiple desirable SPDC processes using a single poling period. The details of the analysis are summarized in Table 1.
It must be noted that in case of TFLN waveguide also, although the poling periods of the SPDC processes (2) and (3) are very close to each other, they still are not the same. In the next section, we further optimize the poling period of the TFLN waveguide and down-converted wavelengths and investigate on the processes which carry exactly the same phase mismatch.

\section{Optimized TFLN waveguide for enabling multiple desired SPDC processes}
\begin{figure} [ht]
\vspace{-1em}
\begin{center}
         {%
    \includegraphics[width= 5 cm , height = 4 cm]{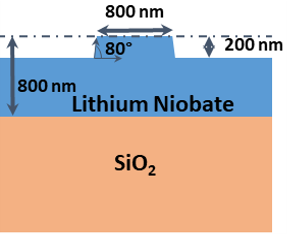}}%
        \caption{ Schematic of the cross-section of TFLN waveguide with top width 800 nm, etching depth 200 nm, thickness 800 nm and sidewall angle 80$^{\circ}$}
        \vspace{-1em}
        \label{FIG. 2}
         \end{center}
        \end{figure} 
As discussed in the previous section, TFLN waveguide due to its strong dispersion gives rise to similar phase mismatches ($\Delta k$) for very different SPDC processes. We optimize and fine-tune the poling period to find the SPDC processes which carry exactly the same phase mismatch. To explore this, we focus on an X-cut TFLN waveguide geometry \cite{arora2024avoiding} with the cross-section shown in Fig. 2. The waveguide has a top width of 800 nm, etching depth of 200 nm, LN thickness of 800 nm, and sidewall angle of  80$^{\circ}$. The dispersion profile of this waveguide has been plotted in Fig. 1 and is represented by dashed (pink and green) lines. It was shown in our earlier works that this waveguide geometry generates spectrally uncorrelated photon pairs at 1550 nm (TM and TE) using pump at 775 nm (TM), and gives rise to heralded single photons with high spectral purity, which remains robust to fabrication imperfections \cite{arora2024avoiding}. The mode structure analysis of this waveguide geometry shows that the waveguide is single-moded for all the interacting wavelengths and  polarizations of the considered SPDC processes. 
\begin{figure*}[b]
        \centering
      \hspace{-1mm}
    {%
    \includegraphics[width= 11 cm,  height = 8 cm]{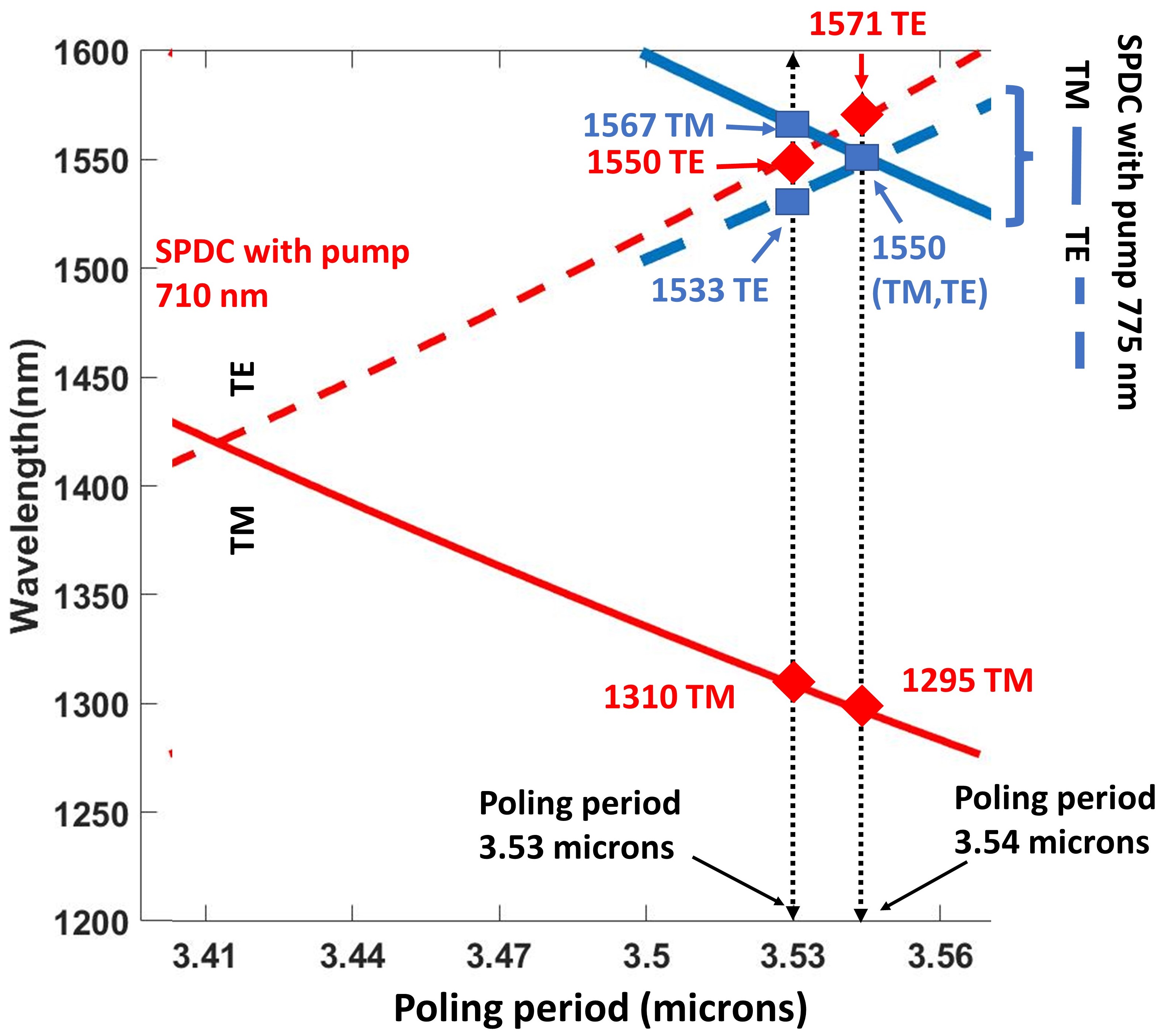}%
  }
  \caption{Variation of the phase-matched wavelengths as a function of poling period for the two SPDC processes, pumped at 710 nm and 775 nm. Red (solid and dashed) lines represent phase-matched signal-idler wavelengths (TM and TE) for the SPDC process pumped at 710 nm; Blue (solid and dashed) lines correspond to phase-matched signal-idler wavelengths (TM and TE) for SPDC process pumped at 775 nm}
   \label{Fig. 3}
\end{figure*}

 \begin{figure*}[ht]
\begin{center}
        {\includegraphics[width=11cm]{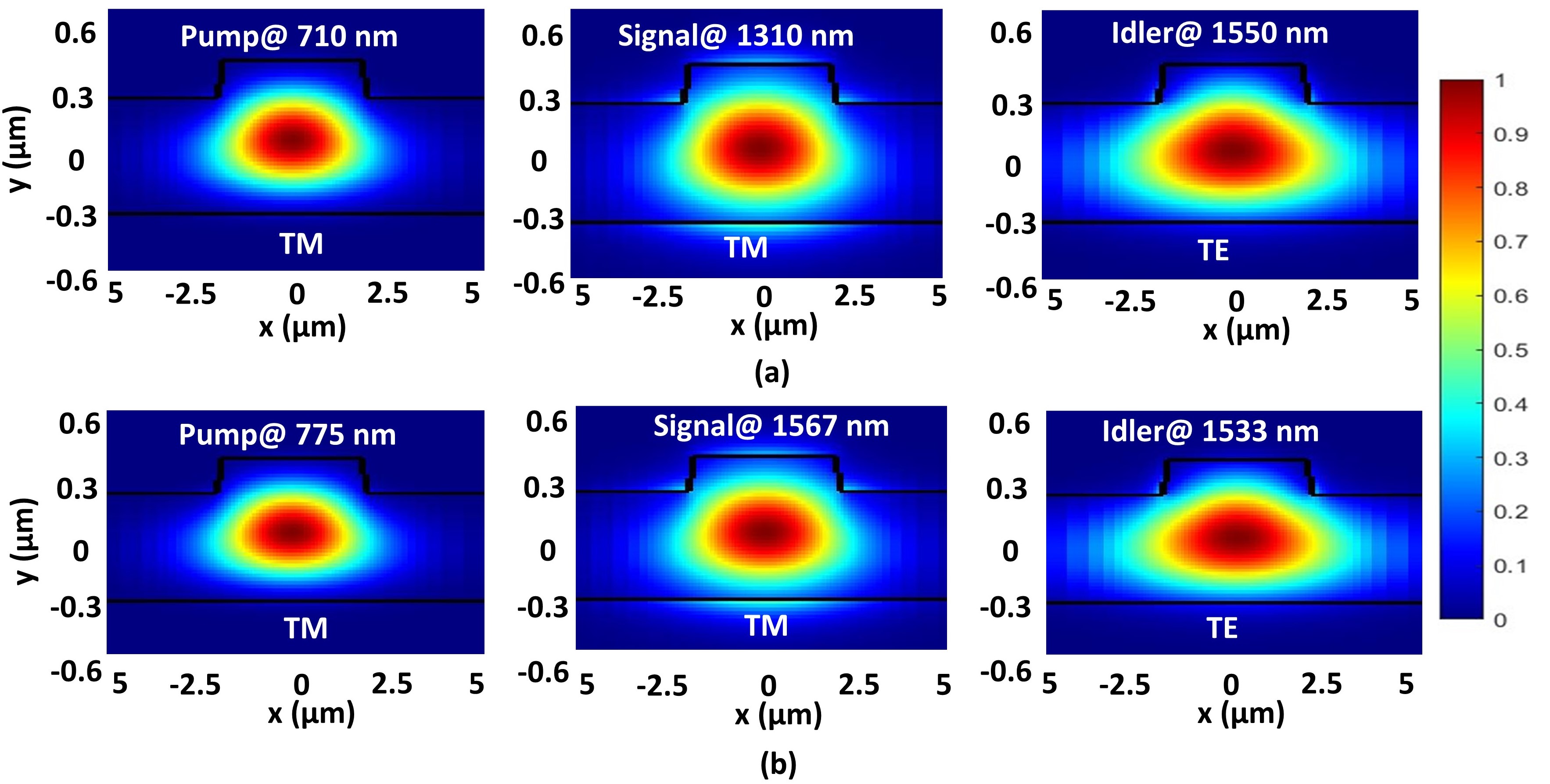 } }
        \caption{Electric field intensity profiles of pump, signal, and idler for the SPDC process pumped at (a) 710 nm and (b) for 775 nm }
        \label{Fig. 4}
        \end{center}
        \end{figure*}

As mentioned in Table 1, for this TFLN waveguide design, the poling periods required to satisfy SPDC processes ((2) and (3)) are 3.54 $\mu m$ and 3.53 $\mu m$ respectively which are very close to each other. So, from here, we seek the possibility of using one common poling period to simultaneously enable two different SPDC processes pumped at 710 nm and 775 nm and giving rise to photon pairs in desired telecom bands. For this, we vary the poling period and compute the phase-matched signal-idler wavelengths which simultaneously satisfy energy and momentum conservation (Eq.2 and Eq. 7) for the two SPDC processes, one pumped at 710 nm and the other at 775 nm. Figure 3 shows the variation of the computed phase-matched signal-idler wavelengths as a function of the poling period. Red (solid and dashed) lines represent phase matched signal-idler wavelengths (TM and TE polarized) for the 710 nm pumped SPDC process, while blue (solid and dashed) lines correspond to signal-idler wavelengths (TM and TE polarized) for the SPDC process pumped at 775 nm. As we can observe from Fig.3, for the poling period of 3.53 $\mu m$ both the SPDC processes generate photon pairs, with their wavelengths lying within the desired range of telecom band (O and C). For this poling period, the first SPDC process (pumped at 710 nm) gives rise to non-degenerate cross-polarized photons at 1310 nm (TM) and 1550 nm (TE), while the second SPDC process (pumped at 775 nm) results in non-degenerate photons within telecom C band at wavelengths 1567 nm (TM)  and 1533 nm (TE). Thus, we show that using this waveguide geometry, a single poling period can enable two different SPDC processes, each giving rise to photon pairs at the desired wavelengths by just tuning the pump wavelength. The corresponding transverse electric field intensity profiles of the interacting photons in the two SPDC processes are shown in Fig. 4.  \\
Once generated, it is also important to study and characterize the properties of the biphoton output state. In the next section, we investigate the spectral properties of the photons generated in the respective SPDC processes.

\section{Spectral properties of biphoton state and Joint spectral intensity}
One of the important features of the generated biphoton state from SPDC is the spectral correlations between the signal and idler photons.  Both spectrally correlated and uncorrelated photon pairs are interesting for quantum optics experiments and have their own respective applications for various tasks\cite{wang2020integrated,nielsen2010quantum,wang2016experimental,knill2001scheme}. In this work, we also study the spectral properties of the generated photon pairs in the respective SPDC processes. For this, we consider the two-photon output state $\psi(\omega_s, \omega_i)$ of SPDC process given by\cite{lugani2020spectrally,jin2013widely}

\begin{equation}
    |\psi(\omega_s, \omega_i)\rangle = \iint f(\omega_s , \omega_i) a^{\dag}_{s} a^{\dag}_{i} |0_s\rangle |0_i\rangle d\omega_s d\omega_i
\end{equation}
Here, $f(\omega_s, \omega_i)$ is the joint spectral amplitude (JSA) which is a product of pump envelope function (PEF), $\alpha(\omega_s + \omega_i)$ and phase matching function (PMF), $\phi(\omega_s, \omega_i)$, \cite{jin2013widely,lugani2020spectrally} where
 \begin{equation}
    \alpha(\omega_s + \omega_i) = exp{\left[-\frac{(\omega_s + \omega_i - \omega_p)^2}{\sigma_p^2}\right]}
    \label{Eq. (2)}
\end{equation}  
  and                            
 \begin{equation}
    \phi(\omega_s, \omega_i) = sinc\left[\Delta k(\omega_s, \omega_i)\frac{L}{2}\right] exp{\left[i\Delta k(\omega_s, \omega_i)\frac{L}{2}\right]}
\end{equation}  \textit{$\omega_{p(s,i)}$}  is the angular frequency of pump (signal, idler) photon, \textit{$\sigma_p$} is the pump bandwidth, \textit{L} is the length of the waveguide and \textit{$\Delta k$} is the phase mismatch.
PEF is governed by the energy conservation within the spectral bandwidth of the pump while PMF is determined by the properties of the non-linear medium. The slope of PMF can be tailored by choosing the appropriate waveguide geometry such that photons with the desired spectral properties can be obtained at the output \cite{kumar2022group,arora2024avoiding}. The slope of the PMF can be described in terms of the group indices of the pump ($n_{g}^{p}$), signal ($n_{g}^{s}$), and idler( $n_{g}^{i}$) modes as\cite{kumar2022group}
\begin{equation}
    tan(\theta) = -\frac{n_{g}^{p} - n_{g}^{s}}{n_{g}^{p} - n_{g}^{i}}
\end{equation}
where $\theta$ is the angle made by the PMF with the horizontal (signal wavelength) axis. 
For spectrally uncorrelated photons,  $\theta$ should satisfy $0^{\circ} \leq \theta \leq 90^{\circ}$, according to which the group index ($n_{g}^{p}$) of the pump photon should either lie between the group indices of signal ($n_{g}^{s}$) and idler ($n_{g}^{i}$) photons or should be equal to one of them  \cite{kumar2022group} i.e, $n_{g}^{s} \leq n_{g}^{p} \leq n_{g}^{i}  \text{ or } n_{g}^{i} \leq n_{g}^{p} \leq n_{g}^{s} $.  This condition is called the group index matching. In Fig. 5 we plot the variation of the group indices of the TM and TE  modes of the considered waveguide design as a function of wavelength.

              \begin{figure}[h]
\begin{center}
        {\includegraphics[width=8cm]{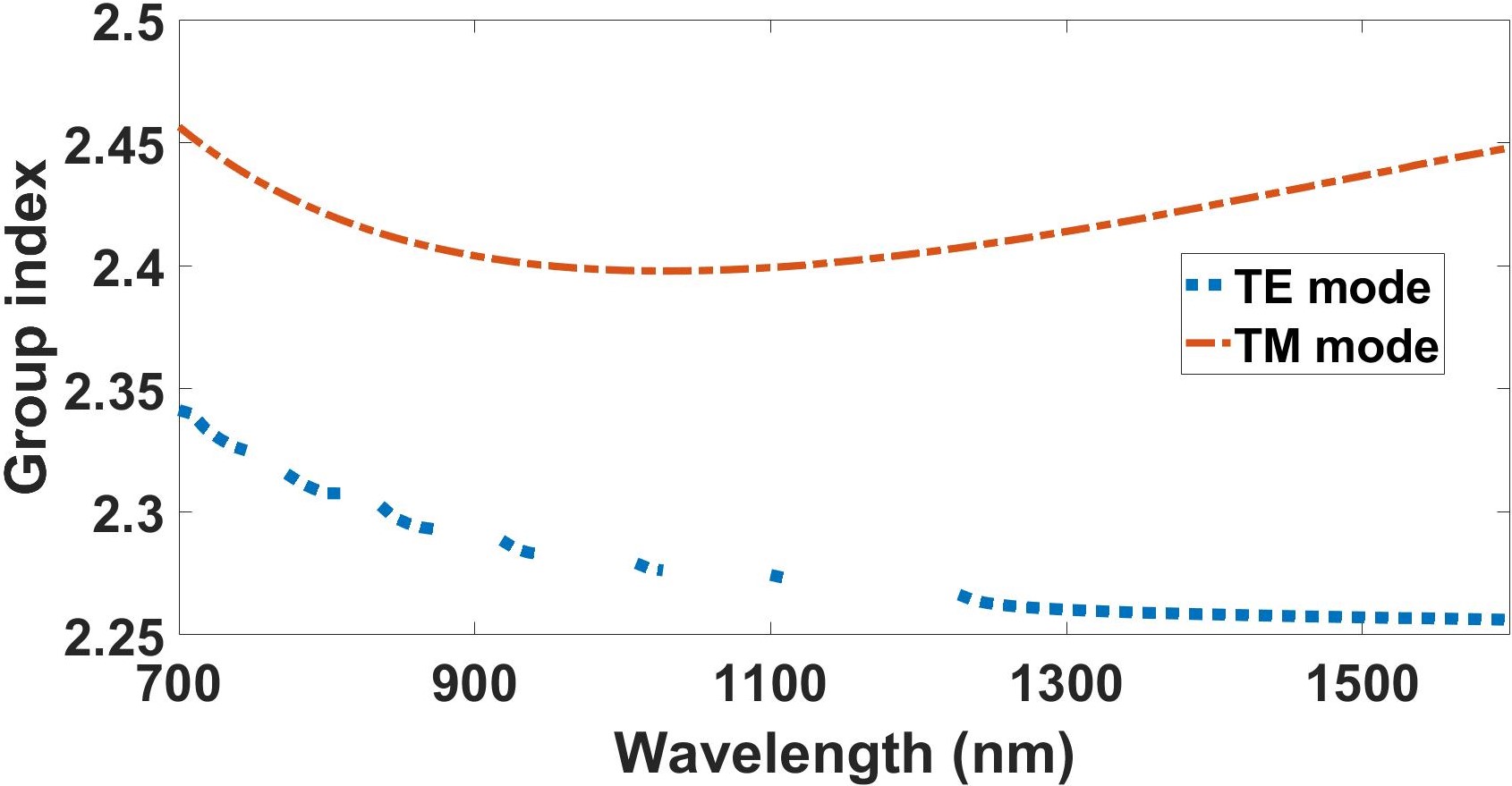 } }
        \caption{ Variation of the group index $n_g$ of fundamental TM and TE modes of the TFLN waveguide as a function of wavelength}
         \vspace{-1.5em}
        \label{Fig. 5}
        \end{center}
        \end{figure} 

     \begin{figure*}[ht]
\begin{center}
        {\includegraphics[width=13cm]{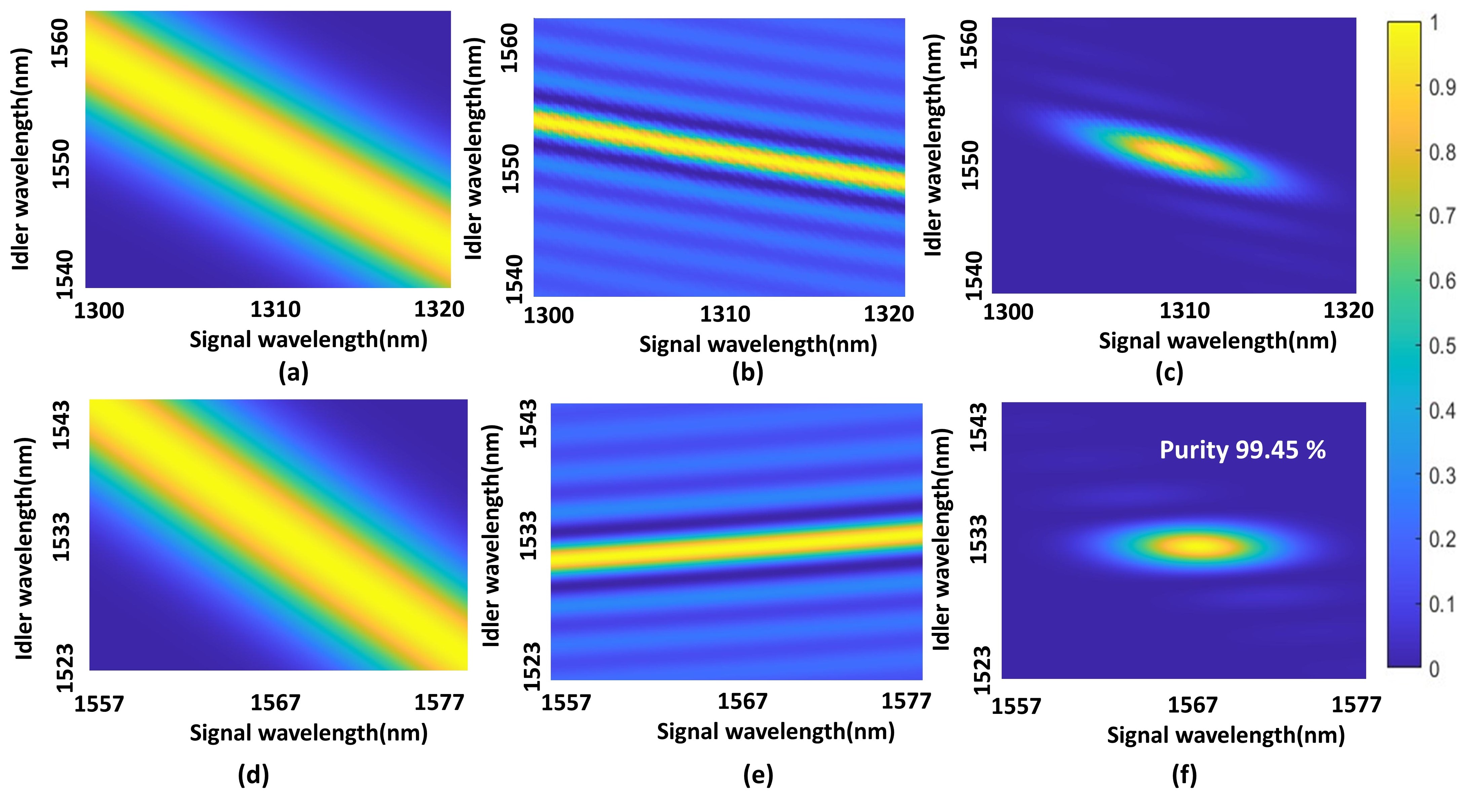 } }
        \caption{(a) Pump envelope function (PEF) (b) Phase matching function (PMF) (c) Joint spectral intensity (JSI) function for the SPDC process pumped at 710 nm ; (d) Pump envelope function (PEF) (e) Phase matching function (PMF) (f) Joint spectral intensity (JSI) function for the SPDC process pumped at 775 nm}
        \label{Fig. 7}
         \end{center}
        \end{figure*}  
As we can see from Fig. 5, that for the SPDC process pumped at 710 nm, the values of group indices (Fig. 5) for pump ($n_{g}^{p}$), signal ($n_{g}^{s}$), and idler ($n_{g}^{i}$) modes come out to be 2.4517, 2.4149 and 2.2567 respectively for the given waveguide geometry. The PMF angle $\theta$ for this process (calculated using Eq. (11)) is found out to be -10.69$^{\circ}$, indicating a spectrally correlated biphoton state. 
 In order to confirm this, we compute the joint spectral intensity (JSI) of the two-photon state ($|f(\omega_s, \omega_i)|^2$).
Fig.6(a-c) shows the PEF, PMF, and JSI functions for this process. The pump bandwidth in these simulations is taken to be 1.5 nm. It can be seen from the slope of the PMF and shape of the JSI of this state that the generated photons are spectrally correlated.
We also calculate the Schmidt number \textit{K} using singular value decomposition (SVD) \cite{mosley2007generation,edamatsu2011photon} of JSI \cite{mosley2007generation}, which quantitatively gives an idea about the correlations between generated signal-idler pairs. For a spectrally uncorrelated state, \textit{K} is equal to 1.  \textit{K} $>$ 1 indicates the presence of correlations in the down-converted photons. For this SPDC process, \textit{K} is found out to be 1.61 which further confirms the presence of the correlations. These correlations can further be optimized by increasing the pump bandwidth. \\
For the second SPDC process (pumped with 775 nm), the values of the group indices (Fig. 5) of pump ($n_{g}^{p}$), signal ($n_{g}^{s}$), and idler ($n_{g}^{i}$) photons are 2.4276, 2.4444, and 2.2568 respectively which satisfy the group index matching condition ($n_{g}^{p}$ $\approx$ $n_{g}^{i}$), required for generating spectrally uncorrelated photon pairs. In this case, the PMF angle '$\theta$' computed from Eq. (11) is 5.6$^{\circ}$. Fig.6(d-f) shows the PEF,  PMF, and JSI for this process. The schmidt number for this process is found to be 1.0056. We calculate the spectral purity of the generated two-photon state using SVD of JSI which is found to be 99.45$\%$. This ensures the generation of heralded single photons with high spectral purity using the type II SPDC process. 

 We would like to mention here that spectrally uncorrelated photons are also possible for telecom O band using this waveguide geometry. For this, one needs to reverse the polarizations of 1310 nm and 1550 nm (in SPDC process 3) photons and the required poling period is found to be 3.29 microns.

\section{Conclusion}
In conclusion, we report on a singly poled TFLN waveguide design that enables completely different and yet desirable SPDC processes giving rise to photon pairs in different telecom bands (O and C) with tunable spectral properties. We focus on two important type II SPDC prcoesses, one process results in spectrally correlated photons at wavelengths 1310 nm (telecom O band) and 1550 nm (telecom C band) while the other gives the spectrally uncorrelated photons in telecom C band at wavelengths 1567 nm and 1533 nm. We attribute this versatility of TFLN to its strong dispersion properties and confirm this by performing a comparitive study between different waveguide systems. The reported design does not require any special domain engineering and relies on a single poling period which is easy to fabricate. Such a versatile source of photon pairs offers enhanced flexibility and would serve as a crucial resource and asset for multiple quantum optics experiments which require quantum states of light at different wavelengths.

\backmatter

\bmhead{Acknowledgments}
This work is supported by a startup research grant by SERB, India (SRG/2022/001759) and PhD scholarships by Ministry of Human Resource Development (MHRD).

\bmhead{Data availability}
The data supporting the findings of this work can be made available from the authors upon reasonable request.

\section*{Declarations}

\bmhead{Conflict of interest} The authors have no conflicts to disclose.
\bmhead{Ethics approval} Not applicable.                        

\bibliography{sn-article}

\end{document}